\documentstyle[11pt,aaspp4,tighten]{article}

\begin{document}

\font\twelvei = cmmi10 scaled\magstep1 
       \font\teni = cmmi10 \font\seveni = cmmi7
\font\mbf = cmmib10 scaled\magstep1
       \font\mbfs = cmmib10 \font\mbfss = cmmib10 scaled 833
\font\msybf = cmbsy10 scaled\magstep1
       \font\msybfs = cmbsy10 \font\msybfss = cmbsy10 scaled 833
\textfont1 = \twelvei
       \scriptfont1 = \twelvei \scriptscriptfont1 = \teni
       \def\mit{\fam1 }
\textfont9 = \mbf
       \scriptfont9 = \mbfs \scriptscriptfont9 = \mbfss
       \def\bmit{\fam9 }
\textfont10 = \msybf
       \scriptfont10 = \msybfs \scriptscriptfont10 = \msybfss
       \def\bmsy{\fam10 }

\def\etal{{\it et al.~}}
\def\eg{{\it e.g.}}
\def\ie{{\it i.e.}}
\def\lsim{\raise0.3ex\hbox{$<$}\kern-0.75em{\lower0.65ex\hbox{$\sim$}}} 
\def\gsim{\raise0.3ex\hbox{$>$}\kern-0.75em{\lower0.65ex\hbox{$\sim$}}} 
 
\title{Estimation of the Mass Outflow Rate From Compact Objects }
 
\author{Sandip K. Chakrabarti\footnote[1]
{Submitted:24th July, 1997; Accepted: ......}}
\affil{S.N. Bose National Centre for Basic Sciences,
JD-Block, Sector III, Salt Lake, Calcutta 700091, India}

\vskip 1cm
\begin{abstract}

Outflows are common in many astrophysical systems which 
contain black holes and neutron stars. Difference between stellar
outflows and outflows from these systems is that the outflows in these
systems have to form out of the inflowing material only. The inflowing material
can form a hot and dense cloud surrounding the compact object either 
because of centrifugal barrier or a denser barrier due to pair plasma or pre-heating effects. 
This barrier behaves like a stellar surface as far as the mass loss is concerned. 
We estimate the outflow rate from such considerations. 
These estimated rates roughly match with the rates in real observations 
as well as those obtained from numerical experiments.

\end{abstract}


\bigskip

Submitted: 24th July, 1997; Revised: 6th November, 1998.
Comparison with Begelman and Blandford work on Cauldron Model is added. 

\clearpage
 
\section{INTRODUCTION}
 
Cosmic radio jets are believed to be originated from the
centers of active galaxies which harbor black holes 
(e.g., Chakrabarti 1996a, hereafter C96a). Even in so called 
`micro-quasars', such as GRS 1915+105 which are believed to have
stellar mass black holes (Mirabel \& Rodriguez, 1994), the outflows are common. 
The well collimated outflows in SS 433 is well known for almost two decades (Margon, 1984).
Similarly, systems with neutron stars also show outflows as is 
believed to be the case in X-ray bursters (e.g. Titarchuk, 1994).

There are large number of works in the literature
which attempt to explain the origin of these outflows. These works 
can be broadly divided into three sets. In one set, the jets
are believed to come out due to hydrodynamic or magneto-hydrodynamic 
pressure effects and are treated separately from the disks 
(e.g., Blandford \& Payne 1982; Fukue 1982; Chakrabarti 1986;
Contopoulos 1995). In another set, efforts are made to 
correlate the disk structure with that of the
outflow (e.g., K\"onigl 1989; Chakrabarti \& 
Bhaskaran 1992). In the third set, numerical simulations are carried out
to actually see how matter is deflected from the equatorial plane 
towards the axis (e.g., Hawley, Smarr \& Wilson 1985; Eggum, 
Coroniti \& Katz, 1985; Molteni, Lanzafame \& Chakrabarti 1994; 
Molteni, Ryu \& Chakrabarti 1996). Nevertheless,
the definitive understanding of the formation of outflows is still
lacking, and more importantly, it has always been difficult to
estimate the outflow rate from first principles. In the first set, 
the outflow is not self-consistently derived from the inflow. 
In the second set, only self-similar steady solutions are 
found and in the third set, either a Keplerian
disk or a constant angular momentum disk was studied, neither being
the best possible assumption. On the other hand, the mass outflow
rate of the normal stars are calculated very accurately from the 
stellar luminosity. Theory of radiatively driven winds seems to be
very well understood (e.g., Castor, Abott  \& Klein, 1975). Given that
the black holes and the neutron stars are much simpler celestial objects, 
and the flow around them is sufficiently hot to be 
generally ionized, it should have been simpler to compute the 
outflow rate from an inflow rate than the methods employed 
in stellar physics.

Our approach to the mass outflow rate computation is somewhat different
from that used in the literature so far. Though we consider simple
minded equations to make our points, such as those applicable 
to conical inflows and outflows, we add a fundamental ingredient 
to the system, whose importance is being revealed only very recently 
in the literature. This is the quasi-spherical centrifugally 
supported dense atmosphere of typical size $5$ to $10$ Schwarzschild 
radius around a black hole and a neutron star. Whether a shock
actually forms or not, this dense region exists, as long as the 
angular momentum of the flow close to the compact object is roughly
constant and is generally away from a Keplerian distribution as 
is the case in reality (Chakrabarti, 1989 [hereafter C89];
Chakrabarti 1996b; hereafter C96b). This centrifugally 
supported region (which basically forms the boundary layer of 
black holes and weakly magnetized neutron stars) successfully
replaced the so called `Compton cloud' (Chakrabarti \& Titarchuk 1995
[hereafter CT95]; Chakrabarti, Titarchuk, Kazanas \& Ebisawa 1996;  
Chakrabarti, 1997 [hereafter C97]) in explaining hard and soft states of black hole
candidates, and the converging flow property of this region successfully
produced the power-law spectral slope in the soft states of black hole
candidates (CT95; Titarchuk, Mastichiadis \& Kylafis 1997).
The oscillation of this region successfully explains the general
properties of the quasi-periodic oscillation (Molteni, Sponholz \& 
Chakrabarti, 1996; Ryu, Chakrabarti \& Molteni 1997) from black holes
and neutron stars. It is therefore of curiosity if this region plays 
any major role in formation of outflows.

Several authors have also mentioned denser regions due to different physical effects.
Chang \& Ostriker, 1985 showed that pre-heating of the gas could  produce standing
shocks at a large distance.  Kazanas \& Ellison (1986) mentioned 
that pressure due to pair plasma could produce standing shocks at smaller distances 
around a black hole as well. Our computation is insensitive to the actual 
mechanism by which the boundary layer is produced. All we require is that the gas should
be hot at the region where the compression takes place. Thus, since Comptonization
processes cool this region (CT95) for larger accretion rates (${\dot M} \gsim 0.1 
{\dot M_{Eddington}}$) our process is valid only for low-luminosity objects, 
consistent with current observations. Begelman \& Rees (1984) talked about 
a so-called `cauldron' model of compact objects where jets were assumed to
emerge from a dense mixture of matter and radiation by boring de-Laval nozzle as in 
Blandford \& Rees (1974) model. The difference between this model and the 
present one is that there very high accretion 
rate was required (${\dot M}_{in} \sim 1000 {\dot M}_E$)  while we consider thermally driven outflows
of smaller accretion rate. Second, the size of the `cauldron' was thousands of
Schwarzschild radii (where gravity was so weak that channel has to have shape of
a de Laval nozzle), while we have a CENBOL of about $10 R_g$ (where the gravity
plays an active role in creating the transonic wind) in our mind. 
Third, in the present case, matter is assumed to pass through a sonic 
point using the pre-determined funnel where rotating pre-jet matter is 
accelerated (Chakrabarti, 1984) and not through a `bored nozzle'
even though symbolically a quasi-spherical CENBOL is considered for mathematical convenience.
Fourth, for the first time we compute the outflow rate completely analytically starting from the
inflow rate alone. To our knowledge such a calculation has not been done in the literature at all.

Once the presence of our centrifugal pressure supported boundary layer (CENBOL) 
is accepted, the mechanism of the formation of the outflow becomes clearer. One basic criteria
is that the outflowing winds should have positive Bernoulli constant (C89). Just as photons
from the stellar surface deposit momentum on the outflowing wind and keeps the flow
roughly isothermal (Tarafdar, 1988) at least upto the sonic point, one may assume 
that the outflowing wind close to the black hole is kept isothermal due to 
deposition of momentum from hard photons. 
In the case of the sun, it's luminosity is only $10^{-5}\ L_{Edd}$ and the typical mass outflow
rate from the solar surface is $10^{-14}M_\odot$ year$^{-1}$ (Priest, 1982). Proportionately, for a
star with a Eddington luminosity, the outflow rate would be $10^{-9} M_\odot$ year$^{-1}$. This is
roughly half as the Eddington rate for a stellar mass star. Thus if the flow is 
compressed and heated at the centrifugal barrier around a black hole, it would also
radiate enough to keep the flow isothermal (at least up to the sonic point) if the efficiency
were exactly identical. Physically, both requirements may be equally
difficult to meet, but in reality with photons shining on outflows near a black hole with almost
$4\pi$ solid angle (from funnel wall) it is easier to maintain the isothermality in the slowly moving
(subsonic) region in the present context. Another reason is  
this: the process of momentum deposition on electrons is more efficient near a black hole.
The electron density $n_e$ falls off as $r^{-3/2}$ while the photon density $n_\gamma$
falls off as $r^{-2}$.Thus the ratio $n_e/n_\gamma \propto r^{1/2}$ increases with the size of the region.
Thus a compact object will have lesser number of electrons per photon and the momentum transfer is
more efficient. In a simpler minded way the physics is scale-invariant, though. In solar physics, it is
customary to chose a momentum deposition term which keeps the flow isothermal to be of the
form (Kopp \& Holzer, 1976],
$$
F_r = \int_{R_s}^r D dr
$$
where, $D$ is the momentum deposition (localized around $r_p$) factor with a typical spatial dependence,
$$
D=D_0 e^{-\alpha (r/r_p-1)^2}
$$ 
Here, $D_0$, $\alpha$ are constants and $R_s$ is the location of the stellar surface.
Since $r$ and $r_p$ comes in ratio, exactly same physical consideration would be
applicable to black hole physics, with the same result {\it provided} $D_0$ is scaled with
luminosity (However, as we showed above, $D_0$ goes up for a compact object.). 
However, as CT95 showed, high accretion rate (${\dot M} \gsim 0.3 {\dot M}_{Edd}$ )
will {\it reduce} the  temperature of the CENBOL catastrophically, and therefore our assumption of
isothermality of the outflow would severely breakdown at these high rates. It is to be noted that
in the context of stellar physics, it is shown (Pauldrachi, Puls \& Kudritzki, 1986) that the temperature
stratification is the outflowing wind has little effect on the mass loss rate. 

Having thus convinced that isothermality of the outflow, at least upto the
sonic point, is easier to maintain near a black hole,
we present in this {\it paper} a simple derivation of the
ratio of the mass outflow rate and mass inflow rate assuming the flow
is externally collimated. We find that the ratio is a function of the compression 
ratio of the gas at the boundary of the hot, dense, centrifugally supported region. 
We estimate that the outflow rate should generally be less than a few percent if the
outflow is well collimated. Finally, in \S 3, we draw our conclusions.

\section{DERIVATION OF THE OUTFLOW RATE}

Assume for the sake of argument that our system is made up of
the infalling gas, the dense boundary layer of the compact object (CENBOL), and 
collimated outflowing wind. Figure 1
shows a schematic diagram of the system with the components marked.
Matter near the equatorial plane is assumed to fall in conical shape onto the
black hole or the neutron star. The sub-Keplerian, 
hot and dense, quasi-spherical region forms either due to centrifugal barrier
or due to pair plasma pressure or pre-heating effects. The outflowing wind 
is assumed to be also conical in shape for simplicity and is flowing out along 
the axis.  It is assumed that the wind is collimated by an external pressure.
Both the inflow and the outflow are assumed to be thin enough so that the velocity and density 
variations across the flow could be ignored.

The accretion rate of the incoming accretion flow is given by,
$$
{\dot M}_{in} = \Theta_{in} \rho \vartheta r^2 .
\eqno{(1)}
$$
Here, $\Theta_{in}$ is the solid angle subtended by the inflow, $\rho$ and
$\vartheta$ are the density and velocity respectively, and $r$ is the
radial distance. For simplicity, we assume geometric units ($G=1=M_{BH}=c$;
$G$ is the gravitational constant, $M_{BH}$ is the mass of the central black hole,
and $c$ is the velocity of light) to measure all the quantities. 
In this unit, for a freely falling gas,
$$
\vartheta (r)= [\frac{1-\Gamma}{r}]^{1/2}
\eqno{(2)}
$$
and
$$
\rho(r) = \frac {{\dot M}_{in}}{\Theta_{in}}(1-\Gamma)^{-1/2} r^{-3/2}
\eqno{(3)}
$$
Here, $\Gamma/r^2$ (with $\Gamma$ assumed to be a 
constant) is the outward force due to radiation. 

We assume that the boundary of the denser cloud is at $r=r_s$
(typically a few Schwarzschild radii, see, Chakrabarti 1996b)
where the inflow gas is compressed. The compression could be
abrupt due to standing shock or gradual as in a shock-free flow
with angular momentum. This details are irrelevant. At this barrier, then 
$$
\rho_+(r_s) = R \rho_- (r_s) 
\eqno{(4a)}
$$
and 
$$
\vartheta_+(r_s) = R^{-1} \vartheta_- (r_s) 
\eqno{(4b)}
$$
where, $R$ is the compression ratio.  Exact value of the compression ratio
is a function of the flow parameters, such as the specific energy and the
angular momentum (e.g., C89; Chakrabarti 1990 [hereafter C90], Chakrabarti, 1996c).
Here, the subscripts $-$ and $+$ denote the pre-shock and post-shock 
quantities respectively. At the shock surface, the total pressure 
(thermal pressure plus ram pressure) is balanced.
$$
P_- (r_s) + \rho_- (r_s) \vartheta_-^2 (r_s)
= P_+ (r_s) + \rho_+ (r_s) \vartheta_+^2 (r_s).
\eqno{(5)}
$$
Assuming that the thermal pressure of the pre-shock incoming flow is 
negligible compared to the ram pressure, using eqs. 4(a-b) we find,
$$
P_+(r_s) = \frac{R-1}{R} \rho_-(r_s) \vartheta_-^2 (r_s).
\eqno{(6)}
$$
The isothermal sound speed in the post-shock region is then,
$$
C_s^2= \frac{P_+}{\rho_+}=\frac{(R-1)(1-\Gamma)}{R^2}\frac{1}{r_s}
=\frac{(1-\Gamma)}{f_0 r_s}
\eqno{(7)}
$$
where, $f_0=R^2/(R-1)$. 
An outflow which is generated from this dense region with very low flow
velocity along the axis is necessarily subsonic in this region,
however, at a large distance, the outflow velocity is expected to be
much higher compared to the sound speed, and therefore the flow must be
supersonic. In the subsonic region of the outflow, the pressure and density
are expected to be almost constant and thus it is customary to 
assume isothermality condition up to the sonic point (Tarafdar, 1988). As argued in the
introduction, in the case of black hole accretion also, such an assumption
may be justified. With isothermality assumption or a given temperature
distribution ($T \propto r^{-\beta}$ with $\beta$ a constant) the result 
is derivable in analytical form. The sonic point conditions are obtained 
from the radial momentum equation, 
$$
\vartheta \frac{d\vartheta}{dr} + \frac{1}{\rho}\frac{dP}{dr} 
+\frac{1-\Gamma}{r^2} = 0 .
\eqno{(8)}
$$
and the continuity equation
$$
\frac{1}{r^2}\frac{d (\rho \vartheta r^2)}{dr} =0
\eqno{(9)}
$$
in the usual way, i.e., by eliminating $d\rho/dr$,
$$
\frac{d\vartheta}{dr}= \frac{N}{D}
\eqno{(10)}
$$
where
$$
N=\frac{2 C_s^2}{r} - \frac{1-\Gamma}{r^2}
$$
and
$$
D=\vartheta - \frac{C_s^2}{\vartheta}
$$
and putting $N=0$ and $D=0$ conditions. These conditions
yield, at the sonic point $r=r_c$, for an isothermal flow,
$$
\vartheta (r_c) = C_s .
\eqno{(11a)}
$$
and
$$
r_c = \frac{1-\Gamma}{2 C_s^2}=\frac {f_0 r_s}{2}
\eqno{(11b)}
$$
where, we have utilized eq. (7) to substitute for $C_s$. 

Since the sonic point of a hotter outflow
is located closer to the black hole, clearly, the condition of isothermality
is best maintained if the temperature is high enough. However if the temperature
is too high, so that $r_c <r_s$, then the flow has to bore a hole through the
cloud just as in the `cauldron' model of Begelman \& Rees (1984), although
it is a different situation --- here the temperature is high, while in the `cauldron' model
the temperature was low. In reality, a pre-defined funnel caused by
centrifugal barrier does not require to bore any nozzle at all,
but our simple quasi-spherical calculation fails to describe this case properly.
This is done in detail in Das \& Chakrabarti (submitted).

The constancy of the integral of the radial momentum equation 
(eq. 8) in an isothermal flow gives: 
$$
C_s^2 ln \ \rho_+ -\frac{1-\Gamma}{r_s} =
\frac{1}{2}C_s^2 + C_s^2 ln \ \rho_c -\frac{1-\Gamma}{r_c}
\eqno{(12)}
$$
where, we have ignored the initial value of the outflowing 
radial velocity $\vartheta (r_s)$ at the dense region boundary, 
and also used eq. (11a). We have also put $\rho(r_c)=\rho_c$ 
and $\rho(r_s) = \rho_+$. Upon simplification, we obtain,
$$
\rho_c =\rho_+  exp (-f)
\eqno{(13)}
$$
where,
$$
f= f_0 - \frac{3}{2}
$$
Thus, the outflow rate is given by,
$$
{\dot M}_{out} = \Theta_{out} \rho_c \vartheta_c r_c^2 
\eqno{(14)}
$$
where, $\Theta_{out}$ is the solid angle subtended by the outflowing cone. 
Upon substitution, one obtains,
$$
\frac{{\dot M}_{out}} {{\dot M}_{in}} = R_{\dot m}
=\frac{\Theta_{out}}{\Theta_{in}} \frac{R}{4} f_0^{3/2} exp \ (-f)
\eqno{(15)}
$$
which, explicitly depends only on the compression ratio:
$$
\frac{{\dot M}_{out}}{{\dot M}_{in}} =R_{\dot m}=
\frac{\Theta_{out}}{\Theta_{in}}\frac{R}{4} 
[\frac{R^2}{R-1}]^{3/2} exp  (\frac{3}{2} - \frac{R^2}{R-1})
\eqno{(16)}
$$
apart from the geometric factors. Notice that this simple result 
does not depend on the location of the sonic points or the
the size of the dense cloud or the outward radiation 
force constant $\Gamma$. This is because the Newtonian potential 
was used throughout and the radiation force was also assumed 
to be very simple minded ($\Gamma/r^2$). Also, 
effects of centrifugal force  was ignored.  Similarly, the ratio
is independent of the mass accretion rate which should be valid only for
low luminosity objects. For high luminosity flows, Comptonization would
cool the dense region completely (CT95) and the mass loss will be negligible.
Pair plasma supported quasi-spherical shocks forms for low luminosity as well
(Kazanas \& Ellison, 1986). In reality there would be a dependence on these 
quantities when full general relativistic considerations of the rotating flows are 
made. Exact and detailed computations using both the transonic inflow
and outflow (where the compression ratio $R$ is also computed self-consistently)
are in progress, and the results would be presented elsewhere (Das \& Chakrabarti, submitted).

Figure 2 contains the basic results. The solid curve shows the
ratio $R_{\dot m}$ as a function of the compression ratio $R$ (plotted from $1$ to $7$),
while the dashed curve shows the same quantity as a function of the
polytropic constant $n=(\gamma-1)^{-1}$ (drawn from $n=3/2$ to $3$), $\gamma$ being the adiabatic
index. The solid curve is drawn for any generic compression ratio and the dashed curve is 
drawn assuming the strong shock limit only: 
$R=(\gamma+1)/(\gamma-1)=2n+1$. In both the curves, $\Theta_{out} \sim
\Theta_{in}$ has been assumed for simplicity. Note that if the compression 
does not take place (namely, if the denser region does not form), then
there is no outflow in this model. Indeed for, $R=1$, the ratio 
$R_{\dot m}$ is zero as expected. Thus the driving force of the outflow is primarily coming from
the hot and compressed region.

In a relativistic inflow or for a radiation dominated inflow, $n=3$ and $\gamma=4/3$. 
In the strong shock limit, the compression ratio is $R=7$ and the ratio 
of inflow and outflow rates becomes,
$$
R_{\dot m}=0.052 \ \frac{\Theta_{out}}{\Theta_{in}}.
\eqno{(17a)}
$$
For the inflow of a mono-atomic ionized gas $n=3/2$ and $\gamma=5/3$. 
The compression ratio is $R=4$, and the ratio in this case becomes,
$$
R_{\dot m}=0.266 \ \frac{\Theta_{out}}{\Theta_{in}}.
\eqno{(17b)}
$$
Since $f_0$ is smaller for $\gamma=5/3$ case, the density at the
sonic point in the outflow is much higher 
(due to exponential dependence of density on $f_0$, see, eq. 7) 
which causes the higher outflow rate, even when the actual jump in density
in the postshock region, the location of the
sonic point and the velocity of the flow at the sonic point are much lower.
It is to be noted that generally for $\gamma >1.5$ shocks are not
expected (C90), but the centrifugal barrier supported dense region
would still exist. As is clear, the entire behavior of the outflow
depends only on the compression ratio, $R$ and the collimating
property of the outflow $\Theta_{out}/\Theta_{in}$.

Outflows are usually concentrated near the axis, while the inflow is near
the equatorial plane. Assuming a half angle of $10^o$ in each case, 
we obtain,
$$
\Theta_{in}= \frac {2 \pi^2}{9}; \ \ \ \ \ \Theta_{out}= \frac {\pi^3}{162}
$$
and 
$$
\frac{\Theta_{out}}{\Theta_{in}} =\frac{\pi}{36} .
\eqno{(18)}
$$
The ratios of the rates for $\gamma=4/3$ and $\gamma=5/3$ are then
$$
R_{\dot m}=0.0045
\eqno{(19a)}
$$
and 
$$
R_{\dot m}= 0.023
\eqno{(19b)}
$$
respectively. Thus, in quasi-spherical systems, 
in the case of strong shock limit, the outflow rate is at the most a couple 
of percent of the inflow. If this assumption is dropped, then for a cold inflow,
the rate could be higher by about fifty percent (see, Fig. 2).

It is to be noted that the above expression for the outflow rate is strictly
valid if the flow could be kept isothermal at least up to the sonic point. In the
event this assumption is dropped the expression for the outflow rate becomes
dependent on several parameters. As an example, we consider a 
polytropic outflow of  same index $\gamma$ but of a 
different entropy function $K$ (We assume the equation of state to be $P=K\rho^\gamma$, with
$\gamma\neq 1$). The expression (11b) would be replaced by,
$$
r_c=\frac{f_0r_s}{2\gamma}
\eqno{(20)}
$$
and eq. (12) would be replaced by,
$$
n a_+^2 - \frac{1-\Gamma}{r_s}=(\frac{1}{2} + n) a_s^2 - \frac{1-\Gamma}{r_c}
\eqno{(21)}
$$
where $n=1/(\gamma-1)$ is the polytropic constant of the flow
and $a_+=(\gamma P_+/\rho_+)^{1/2}$ and $a_c=(\gamma P_c/\rho_c)^{1/2}$ 
are the adiabatic sound speeds at the starting point and 
the sonic point of the outflow. It is easily shown that
a power law temperature fall off of the outflow ($T\propto r^{-\beta}$) would yield
$$
R_{\dot m}= \frac{\Theta_{out}}{\Theta_{in}} (\frac{K_i}{K_o})^n 
(\frac{f_0}{2\gamma})^{\frac{3}{2}-\beta} ,
\eqno{(22)}
$$
where, $K_i$ and $K_o$ are the entropy functions of the inflow and the outflow. This derivation is
strictly valid for a non-isothermal flow. Since $K_i<K_o$, $n>3/2$ and $f_0$,
for ${\Theta_{out}} \sim {\Theta_{in}}$, $R_{\dot m} <<1$ is guaranteed provided $\beta >\frac{3}{2}$,
i.e., if the temperature falls for sufficiently rapidly. For an isothermal flow $\beta=0$ and the rate 
tends to be higher. Note that since $n\sim \infty$ in this case, any small
jump in entropy due to compression will off-balance the the effect of $f_0^{-3/2}$ factor.
Thus $R_{\dot m}$ remains smaller than unity. The first factor decreases with entropy
jump while the second factor increases with the compression ratio ($R$) when $\beta<3/2$.
Thus the solution is still expected to be similar to what is shown in Fig. 2.
Numerical results of the transonic flow  using non-isothermal equation of state are discussed 
in Das \& Chakrabarti (submitted).

\section{CONCLUDING REMARKS}

Although the outflows are common in many astrophysical systems 
which include compact objects such as black holes and neutron
stars, it had been difficult to compute the outflow rates
since these objects do not have any intrinsic atmospheres
and outflowing matter has to be originated from the 
inflow only. We showed in the present paper, that assuming
the formation of a dense region around these objects (as provided by a
centrifugal barrier, for instance), it is possible to obtain the outflow
rate in a compact form with an assumption of isothermality of the outflow at least
up to the sonic point and the ratio thus obtained seems to be quite reasonable. Computation of the
outflow rate with a non-isothermal outflow explicitly depends on several flow parameters.
Our primary goal in this paper was to obtain the rates, and not the
process of collimation. Since observed jets are generally hollow (Begelman, Blandford \& Rees, 1984)
they must be externally supported (either by ambient medium pressure or 
by magnetic hoop stress). This is assumed here for simplicity.
Our assumption of isothermality of the wind till the sonic point is
based on `experience' borrowed from stellar physics. Momentum deposition from the 
hot photons from the dense cloud, or magnetic heating may or may not isothermalize the 
expanding outflow, depending on accretion rates and covering factors. 
However, it is clear that since the 
solid angle at which photons shine on electrons is close to $4\pi$ (as in a 
narrow funnel wall), and since the number of electrons per photon is much
smaller in a compact region, it may be easier to maintain the isothermality
close to a black hole than near a stellar surface.  

The centrifugal pressure supported region that may be present
in presence of angular momentum was found to 
be very useful in  explaining the soft and the hard states 
(CT95; C97), rough agreement with power-law slopes in 
soft states (CT95; Titarchuk, Kylafis \& Mastichiadis, 1997) 
as well as the amplitude and frequency of Quasi-Periodic 
Oscillations (Molteni, Sponholz \& Chakrabarti 1996; Ryu, Chakrabarti \& Molteni 1997)
in black hole candidates. Therefore, our reasonable estimate 
of the outflow rate from these considerations further supports the view that
such regions may be common around compact objects. Particularly interesting
is the fact that since the wind here is thermally driven, the outflow
rate is higher for hotter gas, i.e., when Comptonization is unimportant, that is,
for low accretion rate.  It is obvious that the non-magnetized 
neutron stars should also have the same dense region
we discussed here and all the considerations mentioned here would be
equally applicable. 

It is to be noted that although the existence of outflows are well known, 
their rates are not. The only definite candidate whose outflow rate is known with 
any certainty is probably SS433 whose mass outflow rate was estimated to be
${\dot M}_{out} \gsim 1.6 \times 10^{-6}  f^{-1} n_{13}^{-1} D_5^2 M_{\sun} $ yr$^{-1}$
(Watson et al. 1986), where $f$ is the volume filling factor, $n_{13}$ 
is the electron density $n_e$ in units of $10^{13}$ cm$^{-3}$, $D_5$ 
is the distance of SS433 in units of $5$kpc. Considering a central 
black hole of mass $10M_{\sun}$, the Eddington rate is ${\dot M}_{Ed} \sim 
0.2 \times 10^{-7} M_{\sun} $ yr$^{-1}$ and assuming an efficiency 
of conversion of rest mass into gravitational energy $\eta \sim 0.06$, the 
critical rate would be roughly ${\dot M}_{crit} = {\dot M}_{Ed} / \eta \sim
3.2 \times 10^{-7} M_{\sun} $ yr$^{-1}$. Thus, in order to produce the outflow rate
mentioned above even with our highest possible estimated $R_{\dot m}\sim 0.4$ (see,
Fig. 2), one must have ${\dot M}_{in} \sim 12.5 {\dot M}_{crit}$ which is very high
indeed. One possible reason why the above rate might have been over-estimated
would be that below $10^{12}$cm from the central mass (Watson et al. 1986), $n_{13} >>1 $ 
because of the existence of the dense region at the base of the outflow. 

In numerical simulations the ratio of the outflow and inflow has been computed
in several occasions (Eggum, Coroniti \& Katz, 1985; Molteni, Lanzafame \& Chakrabarti,
1994). Eggum et al. (1985)  found the ratio to be $R_{\dot m} \sim 0.004$ for a 
radiation pressure dominated flow. This is generally comparable with what we found
above (eq. 19a). In Molteni et al. (1994) the centrifugally driven outflowing wind 
generated a ratio of $R_{\dot m}\sim 0.1$. Here, the angular momentum was present 
in both inflow as well as outflow, and the shock was not very strong. Thus,
the result is again comparable with what we find here.

{}

\begin{center}
{\bf FIGURE CAPTIONS}
\end{center}
\begin{description}

\item[Fig.~1] Schematic diagram of inflow and outflow around a compact object. 
Hot, dense region around
the object either due to centrifugal barrier or due to plasma pressure effect
or pre-heating, acts like a `stellar surface' from which the outflowing wind is developed.

\item[Fig.~2] Ratio ${\dot R}_{\dot m}$ of the outflow rate and the inflow rate
as a function of the compression ratio of the gas at the dense region boundary
(solid curve). Also shown in dashed  curve is its variation with the polytropic
constant $n$ in the strong shock limit. Solid angles subtended by the inflow
and the outflow are assumed to be comparable.

\end{description}

\clearpage

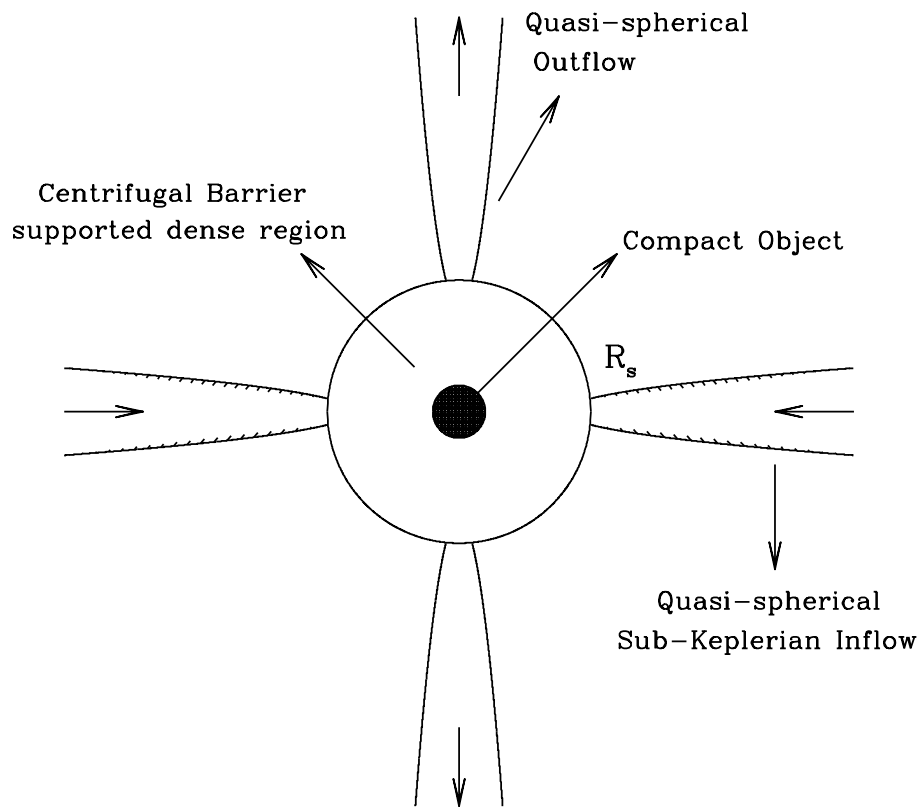
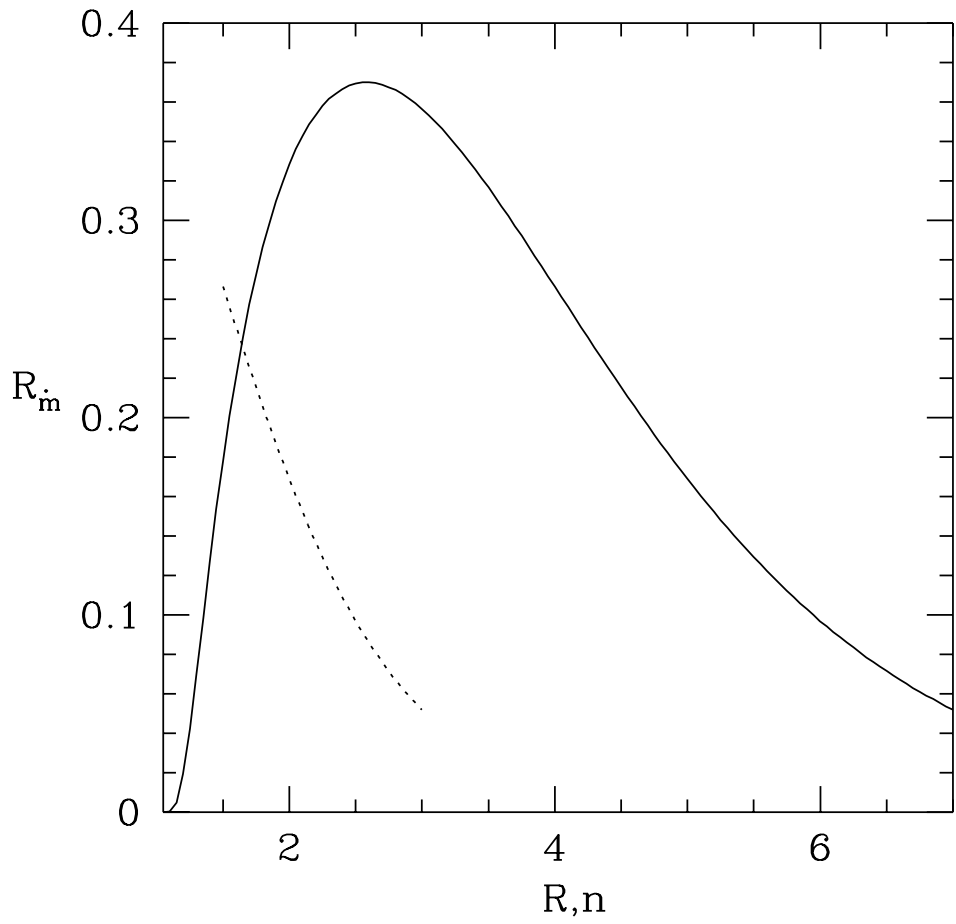
\end{document}